\documentclass[final,1p,times]{elsarticle} 
\usepackage{graphicx} 
\usepackage{amssymb} 
\usepackage{amsthm} 
\usepackage{lineno} 

\journal{Nuclear Physics A} 
\begin{document} 

\begin{frontmatter} 


\title{Monte Carlo for Jet Showers in the Medium}

\author{N\'estor Armesto$^{a}$, Leticia Cunqueiro$^{b}$ and Carlos A. Salgado$^{a}$}

\address[a]{Departamento de F\'\i sica de Part\'\i culas and IGFAE,
Universidade de Santiago de Compostela, E-15706 Santiago de Compostela, Galicia--Spain}

\address[b]{Istituto Nazionale di Fisica Nucleare, Laboratori Nazionali di Frascati,  I-00044 Frascati (Roma), Italy}

\begin{abstract} 
The most commonly employed formalisms of radiative energy loss have been derived in the high-
energy approximation. In its present form, it is reliable only for the medium modifications
of inclusive particle spectra. Modifications to this formalism are expected to be important for less inclusive measurements. This is especially relevant for reconstructed jets in heavy-ion collisions, which are
becoming available only recently. We present some ideas to overcome this limitation. Specifically,
we show an implementation of radiative energy loss within a jet parton shower. This implementation
has been done within the PYTHIA Monte Carlo event generator. We present the publicly available
routine Q-PYTHIA and discuss some of the obtained physics results.
\end{abstract} 

\end{frontmatter} 




\section{Introduction}

Reconstructed jets, a long-awaited probe of medium properties, are starting to become a reality in heavy-ion collisions \cite{Putschke:2008wn,Salur:2008hs}. The expectations at the LHC are very high for these new probes \cite{lhc}. On the theory side, the description of a reconstructed jet as a probe of the medium has already some history. However, analytical approaches are very limited to deal with such a complicated signals --- this is nothing special in heavy-ion collisions but common to other, more simple, systems as well. In this conditions, Monte Carlo codes to describe the jet structure with the needed degree of exclusivity become essential tools for both the phenomenological and the experimental analyses --- where corrections based on these codes are frequent. Several implementations have been presented in the last year, each of them trying to address a particular question of the whole problem. 

On the theoretical level, jets are hard processes which can be described in perturbative QCD. The factorization theorems allow to separate long- from short-distance parts in the cross section. The long distance parts encode all the non-perturbative contributions to the process. An integral part of the cross sections is how these long distance parts evolve with virtuality, which can be computed using perturbative techniques by resumming the relevant splitting probabilities enhanced by phase space factors. The so-called final-state radiation leads to a description of the jet structures. The first and most naive expectation in the case that a medium is formed in heavy-ion collisions is that this radiation pattern is modified. Indeed, most of the successful phenomenology of high-$p_T$ particles at RHIC relied on the {\it radiative energy loss} as the main source of the effect --- see e.g. Ref. \cite{Armesto:2009zi} for a recent study of experimental data. What we present in here is a particular implementation on how this medium-induced radiation can be included in a final-state parton shower and how the corresponding jet structures are modified.

\section{The modified splitting probability}

The gluon radiation spectrum in the presence of a medium, when computed including the important interference with the vacuum contribution, can be split into a medium-induced contribution plus a vacuum contribution \cite{rel}
\begin{equation}
\frac{dI}{d\omega dk_T^2}=\frac{dI^{\rm med}}{d\omega dk_T^2}+\frac{dI^{\rm vac}}{d\omega dk_T^2}
\end{equation}
This factorization has been shown at the level of the one-gluon inclusive distributions. Here, however, we are interested in the {\it exclusive} one-, two-, three- etc gluon distributions.
In the vacuum, the different interference and virtual terms are known to a given accuracy and the dominant contribution for a multi-parton emission is identified to be given by the ordered (in $k_T$ or angle) emission of subsequent gluons, the non-ordered contributions being subleading. A similar computation has not yet been performed for the case of the medium, where, in fact, one would expect that other ordering variables related with the space-time structure of the shower could become relevant --- this space-time structure is irrelevant in the vacuum where no length scale external to the shower is present.

In our implementation, we use a mixed approach, in which the splittings are generated as in the vacuum --- with a similar ordering variable --- but corrections for the space-time evolution are introduced at every splitting. This allows us, in particular, to treat medium- and vacuum-splittings on the same footing \cite{Armesto:2008qh,msf}. In particular, we define a medium-modified splitting probability as
 \begin{equation}
P_{\rm tot} (z)= P_{\rm vac} (z)\to
 P_{\rm tot} (z)=P_{\rm vac}(z)+\Delta P(z,t,\hat{q},L,E).
 \label{eq:splitadd}
 \end{equation}
By matching with the vacuum case, the corresponding modification to the splitting probability is written as
\begin{equation}
\Delta P(z,t,\hat{q},L,E)=\frac{2\pi k_\perp^2}{\alpha_s}\frac{dI^{\rm med}}{d\omega dk_T^2}
\label{eq:medsplit}
\end{equation}
where  $dI^{\rm med}/d\omega dk_T^2$ is given by the medium-induced gluon radiation used previously for RHIC phenomenology \cite{rel}. This spectrum depends on two parameters, the medium length $L$ and the transport coefficient $\hat q$. The latter of these quantities encodes all possible information about the medium properties as temperature, density, etc. in a single parameter, with the interpretation of the average transverse momentum squared that the gluon gets per mean free path in the medium. 

\section{Implementation and results}

The medium-modified splitting probabilities (\ref{eq:splitadd}, \ref{eq:medsplit}) are implemented in a modification of the standard PYTHIA \cite{Sjostrand:2006za} routine for final-state parton branching {\tt PYSHOW}. There, the vacuum splitting functions are supplemented with the medium term --- see \cite{Armesto:2008qh} for details. The modified routines along with auxiliary routines to compute the medium contribution to the splitting probability and geometry are released for public use and available at the Site  \cite{qatmc}. As a short name, we call this implementation Q-PYTHIA \cite{Armesto:2008qh}.

\begin{figure}
\begin{center}
\begin{minipage}{0.54\textwidth}
\includegraphics[width=\textwidth]{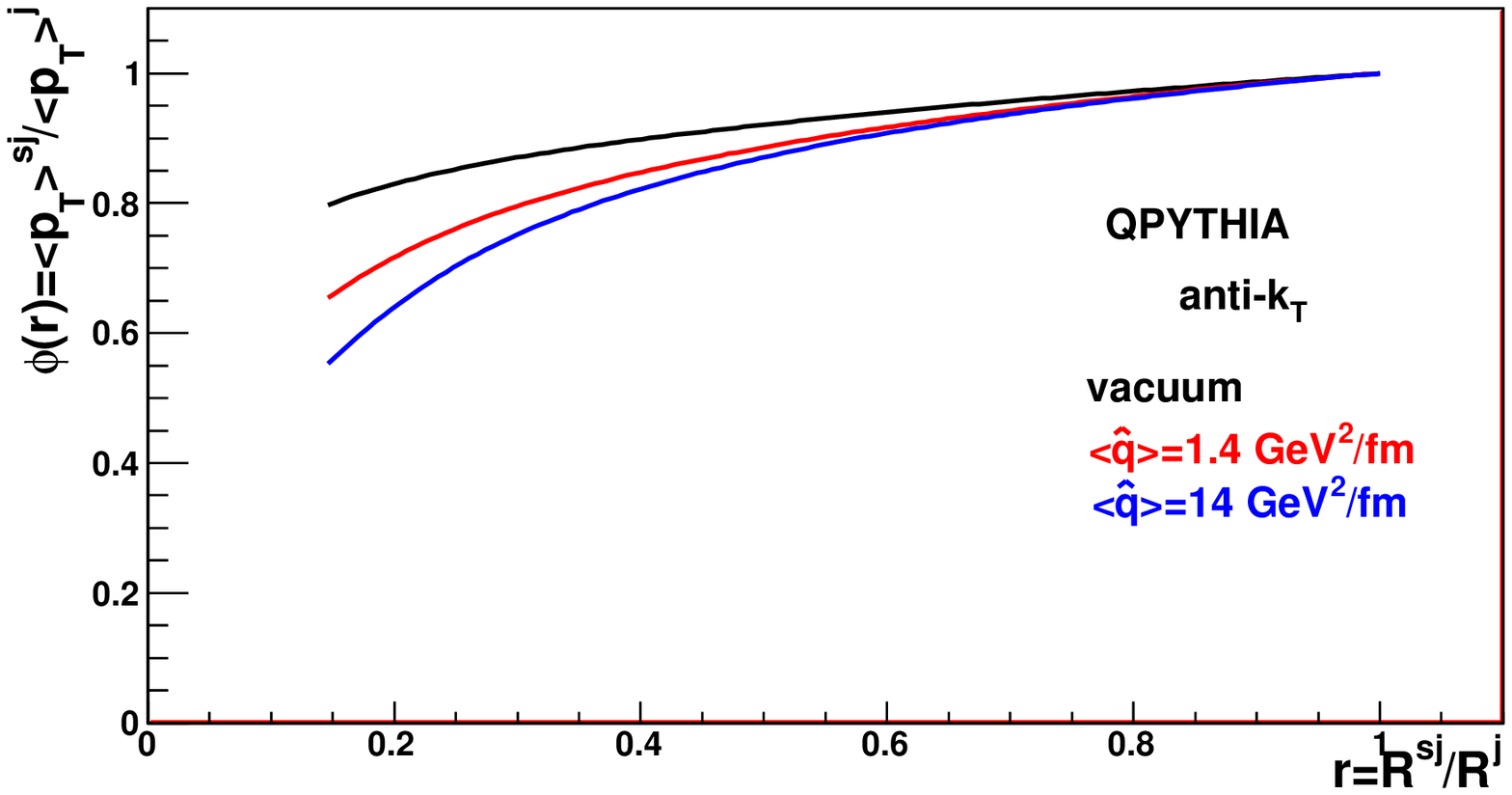}
\end{minipage}    
\hfill
\begin{minipage}{0.45\textwidth}
    \includegraphics[width=\textwidth]{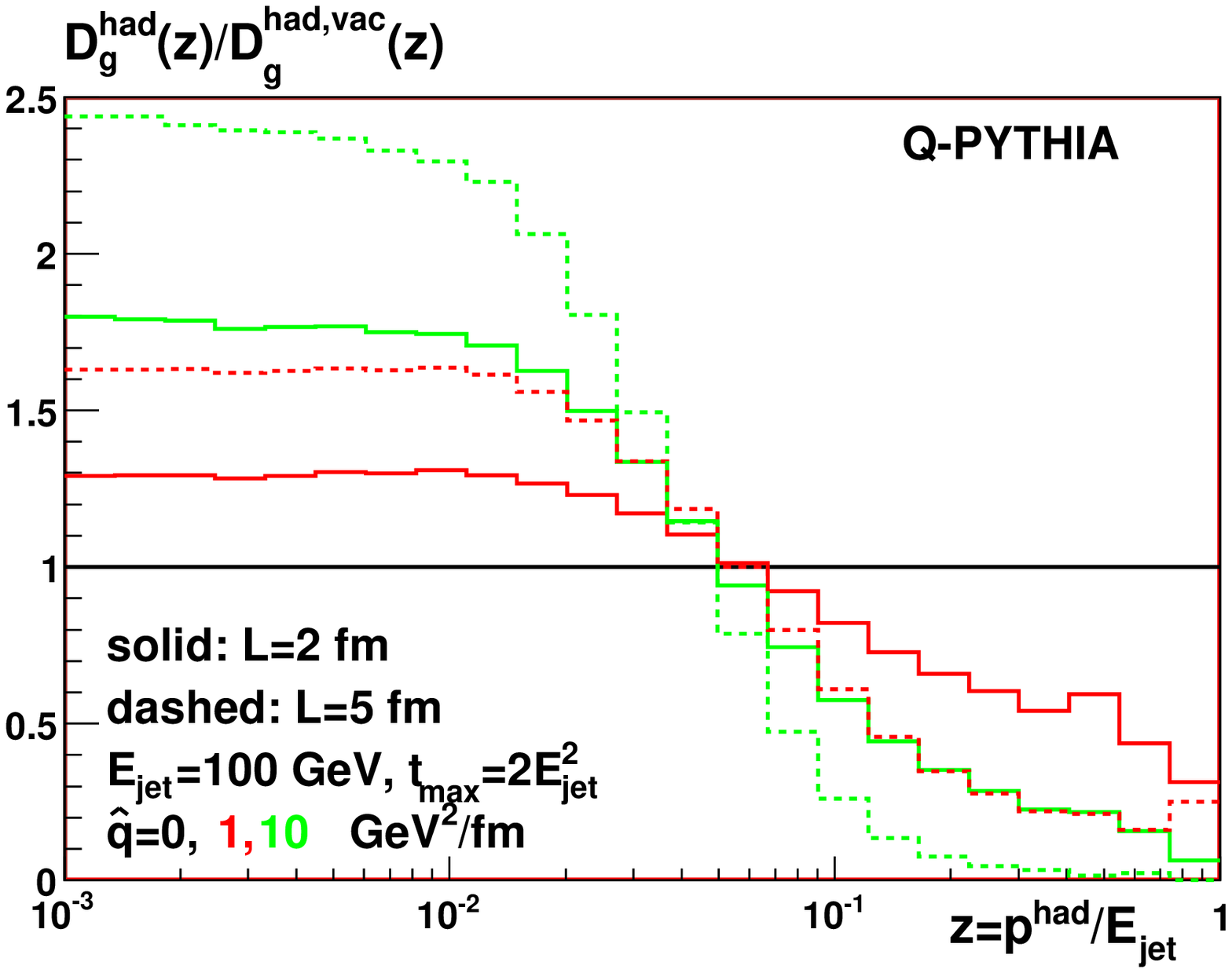}
\end{minipage}    
\end{center}
\caption{(Left) Jet shapes in pp collisions at $\sqrt{s_{NN}}=5.5$ TeV using the anti-$k_T$ recombination 
algorithm for different quenching parameters. (Right)
Ratio of medium to vacuum fragmentation functions for all hadrons, for different gluon energies $E_{jetÊ}=100$ GeV. Figures from Ref. \protect\cite{Armesto:2008qh}}
\label{fig:1}       
\end{figure}

The additional medium-induced splitting probability is expected to produce:
\begin{enumerate}
\item Softening of the spectra, leading to an apparent energy loss of the leading particle in the jet
\item Increase of the intra-jet multiplicity
\item Broadening of the jet angular structures due to the larger typical angle of radiation in the medium when compared to the vacuum.
\end{enumerate}
These features are observed in the physics results from our implementation Q-PYTHIA  \cite{Armesto:2008qh}. In Fig. \ref{fig:1} the angular energy broadening and the softening of the fragmentation function is plotted for different geometries and values of the transport coefficient. It is worth noticing that a very large energy broadening could induce a bias in the jet reconstruction while (due to the softening of the spectrum) the study of intra-jet particle angular distributions is a good observable for medium effects \cite{Salgado:2003rv}.

\section{Usage of the routines}

Q-PYTHIA is standard PYTHIA with a modified final-state parton shower routine. So, all the features of this Monte Carlo are unchanged except for the evolution of the jet after the hard scattering --- we refer to the standard manual \cite{Sjostrand:2006za} for usage. 

The medium-modifications are introduced in Q-PYTHIA through two {\it geometry} routines\footnote{{\it Geometry} here is to be understood in a generic way as the space-time extension of a medium, including expansion, etc. In fact, what is needed is a transport coefficient defined locally at every point of the space and every time after the collision.}: {\tt QPYGIN} to define the initial position of the hard scattering and {\tt QPYGEO} which returns the medium properties that a parton traversing the medium will measure. In order to allow for a flexible implementation, the geometry of the medium is not fixed but can be defined by the user --- examples of simple geometries are provided and more of them will be available in the future. The medium enters in the splitting probabilities through only two variables, the transport coefficient, $\hat q$, and the in-medium path length, $L$. In fact, the actual variables used in the implementation are a combination of these two: $\hat qL$ and $\omega_c=\hat qL^2/2$. For a generic medium in which the transport coefficient is defined locally $\hat q(x,y,z,\tau)$ one usually defines these quantities as the effective integrals
\begin{eqnarray}
\omega_c^{eff}(x,y,z,\tau,\beta_x,\beta_y,\beta_z)=\int d\xi \, \xi \hat{q}(\xi),\nonumber\\ 
\protect[ \hat{q}L ]^{eff}(x,y,z,\tau,\beta_x,\beta_y,\beta_z)=\int d\xi \, \hat{q}(\xi),
\label{eq:scaling}
\end{eqnarray}
where $x$, $y$, $z$ and $\tau$ are the initial position and time to be considered (the initial production point or the point of last splitting) and $\beta_i$, $i=x,y,z$, the components of the corresponding trajectory three-velocity in units of $c$. The integrations in Eq. (\ref{eq:scaling}) are to be performed along the linear trajectory defined by these inputs and parametrized by $\xi$.

Once the initial point of the jet evolution is fixed by {\tt QPYGIN(X0,Y0,Z0,T0)}, Q-PYTHIA will start the jet evolution by computing the probability of splitting and assigning it a space-time position computed by the formation time of the secondary parton, estimated as
\begin{equation}
t_{\rm form}=\frac{2\omega}{k_T^2}
\label{eq:formtime}
\end{equation}
This time defines then the subsequent position $(x,y,z,\tau)$ where the next splitting will start --- Q-PYTHIA calls then {\tt QPYGEO(X,Y,Z,T,BX,BY,BZ,QHL,OC)}\footnote{When the splitting is the first one, $(x,y,z,\tau)$ will correspond to the initial values $(x_0,y_0,z_0,\tau_0)$.} which returns the values of $\protect[ \hat{q}L ]^{eff}$ and $\omega_c^{eff}$ for this particular path. In this way, the formation time is introduced at every splitting allowing to account for the space-time evolution of the jet structure. So, the standard way of implementing a geometry defined in the routines {\tt QPYGIN} and {\tt QPYGEO} simplifies to: i) defining a local transport coefficient $\hat q(x,y,z,\tau)$ --- for example by assuming $\hat q\propto \epsilon^{3/4}$ and $\epsilon$ the energy density field computed in a hydrodynamical simulation; ii) computing the linear integrals (\ref{eq:scaling}) for each path defined by the inputs $(x,y,z,\tau)$ and $(\beta_x,\beta_y,\beta_z)$.


\section*{Acknowledgments} 
This work has
been supported by Ministerio de Ciencia e Innovaci\'on of Spain under
projects FPA2005-01963, FPA2008-01177 and contracts Ram\'on y Cajal
(NA and CAS); by Xunta de Galicia (Conseller\'{\i}a de Educaci\'on and
Conseller\'\i a de Innovaci\'on e Industria -- Programa Incite) (NA
and CAS); by the Spanish Consolider-Ingenio 2010 Programme CPAN
(CSD2007-00042) (NA and CAS); by the European Commission grant
PERG02-GA-2007-224770 (CAS);

\end{document}